\documentclass[letterpaper, aps, showpacs,onecolumn]{revtex4}

\usepackage{amsmath}
\usepackage{amssymb}
\usepackage{latexsym}
\usepackage{graphicx,epstopdf}
\usepackage{hyperref}

\begin{document}

\title{Brownian motion approach to anomalous rotation of galaxies}
\author{Alexander Jurisch}
\affiliation{ajurisch@ymail.com, Munich, Germany}

\begin{abstract}
It has been shown that the weak-interacting limit of the metric-skew-tensor-gravity (MSTG) can explain the anomalous rotation of galaxies without non-baryonic dark matter. We show that MSTG is related to the equilibrium-state of ordinary Brownian motion. We also explore if other stochastic processes can model anomalous rotation. Furthermore, we analyze phase-diagrams that elucidate the condensation of a gravitating cloud towards a Kepler-Newton system and illustrate regions of existence of rotating objects.
\end{abstract}
\maketitle

\section{Author's note}
The author revokes all changes made in version v2 of this paper, they came from a confusion of angular velocity and path-velocity. Version v1 is correct in all points and identical with the present version v3\,. The author apologizes to the reader.

\section{Introduction}
Anomalous rotation was first discussed by Oort and Zwicky around 1930 and came back when Rubin et. al. \cite{Rubin1, Rubin2}  published their results in 1970\,. Contrary to what was to be expected, the rotation-curves of galaxies heavily deviate from the Kepler-Newton law. Instead to decay like $ v(r)\sim r^{-1/2}$, it was found that the tail of $v(r)$ is flat or even rises like $v(r)\sim r^{n}$, $n\geq 0$\,. It is widely accepted, but still unproven, that anomalous rotation is a consequence of $\Lambda$-cold-dark-matter ($\Lambda$CDM), where $\Lambda$ is the cosmological constant that accounts for dark energy. The $\Lambda$CDM model was first proposed by Ostriker and Steinhardt \cite{Ostriker}.

Approaches that avoid dark matter hypotheses are rare. The most prominent one amongst them is Milgrom's  phenomenological, non-self-contained ad-hoc construction of a modified Newtonian dynamics (MOND), \cite{Milgrom1, Milgrom2, Milgrom3}. MOND is able to reproduce the flatness of the tail of the rotational curve by $v(r\rightarrow\infty)=(M G a_{0})^{1/4}$. Thereby $a_{0}$ is a hypothetic acceleration of unclear origin, that only takes effect in deep space. The MOND phenomenology has found considerable interest, since for a long time it has been the only alternative to $\Lambda$CDM-theories. The literature is vast, and we shall only quote some selected work on this field. Hossenfelder et. al. \cite{Hossenfelder} recently successfully related MOND to Verlinde's Covariant Emergent Gravity (CEG) \cite{Verlinde}. Relativistic generalizations of MOND have been worked out by Bekenstein \cite{Bekenstein} and Sanders \cite{Sanders}. Deffayet et. al. \cite{Deffayet} suggested a nonlocal alternative to Bekenstein's approach. All parent-theories that attempt to derive MOND in the non-relativistic limit but share one serious drawback in the sense that they are ex post constructions and not fundamental.

The price to pay for MOND, however, is that it leads to a breakdown of Newton's second law in deep and empty space. Newton's laws but are made for empty space, and Occam's razor tells us that MOND is thus likely be an epicycle, since a breakdown of Newtonian dynamics is too high a price to pay in order to avoid the still elusive dark matter. By Occam's razor we may conclude that dark matter comes at cheaper cost than MOND. Attempts to explain MOND by dark matter itself have been suggested by e.g. Kaplinghat et. al. \cite{Kaplinghat} and Blanchet et. al. \cite{Blanchet1, Blanchet2}.

Moffat has developed a relativistic non-symmetrical theory of gravity (NTG) \cite{Moffat1}, and furthermore a metric-skew-tensor-gravity (MSTG) \cite{Moffat2} which both are free of dark matter hypotheses. Contrary to MOND and it's parent-theories NTG and MSTG are self-contained and able to explain the anomalous rotation by a modification of the acceleration, see \cite{Brownstein1, Brownstein2}. This modification does not violate Newton's axioms. In the weak-field limit Moffat's theory leads to an Ornstein-Zernike-type or likewise a Yukawa-type correction to the gravitational potential. MSTG is built upon sound foundations, and thus, by Occam's razor, is certainly much more likely to correctly describe the rotational anomaly of galaxies than MOND or dark matter.

The starting point of our considerations here is that a galaxy can be understood as a classical condensed (or still condensing) weakly-interacting many-body system and thus qualifies to be treated by Brownian motion and diffusion. In this spirit we demonstrate that the phenomenon of anomalous rotation can soundly be described by the equilibrium state of a stochastic process. First, we show that the weak-interacting limit of MSTG can be derived from ordinary Brownian motion with spherical symmetry. We also explore if other Brownian processes can model anomalous rotation successfully. To do so, we analyze the well-known Ornstein-Uhlenbeck process and a generalized Brownian process with a power-law drift-field.

Furthermore, we analyze phase-diagrams that connect fundamental parameters like the total angular momentum, the core-radius, the range and the mass-scale of a rotating object. From the phase-diagrams we can make statements about the condensation of a gravitating cloud towards a Kepler-Newton system and identify regions of existence. In the framework of the model the regions of existence can help to determine the initial state of a galaxy. We assume that all our considerations may also hold for proto-planetary disks.

We did not perform an analysis of data, since this work has already been done by Brownstein and Moffat in \cite{Brownstein1, Brownstein2, Moffat2}. There is no reason to assume that the other types of Brownian motion we consider here would not yield a good match on data. Only the parameters that enter the modified gravitational potential would differ.

\section{Mass-model}
The choice for the model of the mass is necessarily somewhat arbitrary. We decided to work with the spherical mass-model that was suggested in \cite{Brownstein1},
\begin{equation}
M(r)\,=\,M\,\left(\frac{r}{R_{\rm{c}}\,+\,r}\right)^{3\,\beta}\quad,
\label{diffusionmass1}\end{equation}
where $R_{\rm{c}}$ denotes the radius of the core and $\beta$ describes different types of phenomenology.
The value $\beta=1$ fits high-surface-brightness (HSB) galaxies, while $\beta=2$ fits low-surface-brightness (LSB) galaxies and dwarf galaxies.
The rotational curve is calculated by equating the gravitational acceleration with the centripetal acceleration
\begin{equation}
\frac{v^{2}}{r}\,=\,M(r)\,\partial_{r}\,\phi(r)\,\rightarrow\,v(r)\,=\,\sqrt{r\,M(r)\,\partial_{r}\,\phi(r)}\quad,
\label{diffusionmass3}\end{equation}
where $\phi(r)$ ist the gravitational potential.

\section{Metric-skew-tensor-gravity by Brownian motion}
In general, the equation of motion of a spatial Brownian process is given by
\begin{equation}
d\mathbf{r}(t)\,=\,\mathbf{a}(\mathbf{r},\,t)\,dt\,+\,\sqrt{D(\mathbf{r},\,t)}\,d\mathbf{W}(t)\quad,
\label{GeneralBrownian1}\end{equation}
where $\mathbf{a}$ is a classical drift-field, $D$ the diffusive function and $d\mathbf{W}$ a spatial Wiener-process. By It\^o's theorem Eq. (\ref{GeneralBrownian1}) is related to the Fokker-Planck equation
\begin{equation}
\partial_{t}\,P(\mathbf{r},\,t)\,=\,-\,\nabla\cdot\left[\mathbf{a}(\mathbf{r},\,t)\,P(\mathbf{r},\,t)\right]\,+\,\nabla^{2}\left[D(\mathbf{r},\,t)\,P(\mathbf{r},\,t)\right]\quad.
\label{GeneralBrownian2}\end{equation}
If in Eq. (\ref{GeneralBrownian2}) $\partial_{t}\,P\,=\,0$ holds, the Brownian process is in it's equilibrium.

The function $P$ can be interpreted as a probability-distribution, however, as Eq. (\ref{GeneralBrownian2}) is a consequence of the transport-theorem we may also interpret $P$ as a potential-function $\phi\,=\,P$ if a source $\rho$ is present,
\begin{equation}
\partial_{t}\,\phi(\mathbf{r},\,t)\,+\,\nabla\cdot\left[\mathbf{a}(\mathbf{r},\,t)\,\phi(\mathbf{r},\,t)\right]\,-\,\nabla^{2}\left[D(\mathbf{r},\,t)\,\phi(\mathbf{r},\,t)\right]\,=\,\rho(\mathbf{r},\,t)\quad.
\label{GeneralBrownian3}\end{equation}

\subsection{Ordinary Brownian Motion}
Here we show that the weak-field regime of MSTG as calculated in \cite{Moffat2} is identical to a Brownian process. To do so, we consider the radial stochastic process
\begin{equation}
dr_{{\rm{B}}}(t)\,=\,-\,\gamma\,dt\,+\,\sqrt{D}\,dW(t)\quad,
\label{MSTG1}\end{equation}
where $dr_{{\rm{B}}}(t)$ denotes the Brownian correction to the classical trajectory, $dW(t)$ is a Gaussian Wiener-process, $D$ the diffusion-constant and $\gamma$ a drift-parameter that induces skewness in the sense that a test-particle tends to move into the direction of the drift. The sign of $\gamma$ has to be negative, since else no physically meaningful potential can be defined. The parameter $\gamma$ has the dimension of a velocity, and we observe that the classical part of the Brownian process, $\left<r_{{\rm{B}}}(t)\right>=-\,\gamma\,t$, generates a drift towards the center of gravity that may be understood as a confinement.
The Fokker-Planck equation of the Brownian process Eq. (\ref{MSTG1}) is given by
\begin{equation}
\partial_{t}\,\phi_{{\rm{B}}}(r,\,t)\,=\,\gamma\,\partial_{r}\,\phi_{{\rm{B}}}(r,\,t)\,+\,D\,\frac{1}{r^{2}}\partial_{r}\,r^{2}\,\partial_{r}\,\phi_{{\rm{B}}}(r,\,t)\,+\,\rho\quad.
\label{MSTG4}\end{equation}
The equilibrium, thus time-independent solution of Eq. (\ref{MSTG4}), $\partial_{t}\phi=0$, follows by
\begin{equation}
\phi_{{\rm{B}}}(r)\,=\,\sigma\left(\frac{\exp\left[-\,r/r_{0}\right]}{r}\,+\,\frac{{\rm{Ei}}\left[-\,r/r_{0}\right]}{r_{0}}\right)\,,\quad r_{0}\,=\,\frac{D}{\gamma}\quad,
\label{MSTG5}\end{equation}
where ${\rm{Ei}}$ the exponential integral and $\sigma$ is the coupling generated by the source $\rho$. The parameter $r_{0}$ defines a correlation-length that describes the effective range of the potential.

By rewriting Eq. (\ref{MSTG5}) we obtain
\begin{equation}
\phi_{{\rm{B}}}(r)\,=\,\sigma\,\frac{\exp\left[-\,r/r_{0}\right]}{r}\,\left(1\,+\,\frac{r}{r_{0}}\,{\rm{Ei}}\left[-\,r/r_{0}\right]\,\exp\left[r/r_{0}\right]\right)\quad.
\label{MSTG6}\end{equation}
The potential Eq. (\ref{MSTG6}) is identical to the general result that was found in \cite{Moffat2} by MSTG. Observing that
\begin{equation}
\lim_{r/r_{0}\rightarrow\,0}\,\frac{r}{r_{0}}\,{\rm{Ei}}\left[-\,r/r_{0}\right]\,\exp\left[r/r_{0}\right]\,=\,0\quad,
\label{MSTG7}\end{equation}
the correction to Newtonian gravity becomes
\begin{equation}
\phi(r)\,=\,\phi_{{\rm{N}}}(r)\,+\,\phi_{{\rm{B}}}(r)\,=\,-\,\frac{G\,M}{r}\,+\,\sigma\,\frac{\exp\left[-\,r/r_{0}\right]}{r}\quad,
\label{MSTG8}\end{equation}
where $\sigma$ is a coupling-constant. For $r\rightarrow 0$ the gravitational potential is
\begin{equation}
\phi(r)\,=\,-\,\frac{G\,M}{r}\,+\,\frac{\sigma}{r}\quad.
\label{MSTG9}\end{equation}
Following \cite{Moffat2} we require that the near-field potential Eq. (\ref{MSTG9}) is equal to Newtonian gravity. This can be taken care of by the ansatz $G_{{\rm{N}}}\,M\,=\,G_{{\rm{d}}}\,M\,-\,\sigma$, where $G_{{\rm{N}}}$ is Newton's constant of gravity and $G_{{\rm{d}}}$ is a dressed (renormalized) constant of gravity. Without loss of generality we may assume that $G_{{\rm{d}}}\,=\,G_{{\rm{N}}}\left(1\,+\,g\right)$. From this follows $\sigma\,=\,g\,G_{{\rm{N}}}\,M$. We may understand $g$ as a fitting-parameter. However, Moffat in \cite{Moffat2} takes this one step further and introduces $g\,=\,\sqrt{M_{0}/M}$, and thus $\sigma\,=\,G_{{\rm{N}}}\,\sqrt{M\,M_{0}}$. The mass-parameter $M_{0}$ is to be understood as a reference-scale of mass. The potential is thus given by
\begin{equation}
\phi(r)\,=\,-\,\frac{G_{{\rm{N}}}\,M\,+\,G_{{\rm{N}}}\,\sqrt{M\,M_{0}}}{r}\,+\,G_{{\rm{N}}}\,\sqrt{M\,M_{0}}\,\frac{\exp\left[-\,r/r_{0}\right]}{r}\quad.
\label{MSTG10}\end{equation}

\subsection{Ornstein-Uhlenbeck process}
A well-known stochastic process is the Ornstein-Uhlenbeck process. We now shall examine if this process can model anomalous rotation. The Ornstein-Uhlenbeck process is defined by
\begin{equation}
dr_{{\rm{OU}}}(t)\,=\,-\,\gamma\,r_{{\rm{OU}}}(t)\,dt\,+\,\sqrt{D}\,dW(t)\quad.
\label{OU1}\end{equation}
Note that the classical trajectory is given by $\left<r_{{\rm{OU}}}(t)\right>\,=\,R\,\exp[-\,\gamma\,t]$. Contrary to the ordinary Brownian process above the classical trajectory relaxes like an exponential.

The Fokker-Planck equation is now given by
\begin{equation}
\gamma\,\partial_{r}\,r\,\phi_{{\rm{OU}}}(r,\,t)\,+\,D\,\frac{1}{r^{2}}\partial_{r}\,r^{2}\,\partial_{r}\,\phi_{{\rm{OU}}}(r,\,t)+\,\rho\,=\,0\quad.
\label{OU2}\end{equation}
The potential can be calculated analytically, reading
\begin{equation}
\phi_{{\rm{OU}}}(r)\,=\,\frac{\sigma}{r}\,\left(1\,+\,\exp\left[-\,\frac{r^{2}}{2\,r_{0}^{2}}\right]\,H_{-1}\left[\frac{r}{\sqrt{2}\,r_{0}}\right]\right),\,r_{0}\,=\,\sqrt{\frac{D}{\gamma}}\quad,
\label{OU3}\end{equation}
where $H_{-1}$ is a Hermite-polynomial. The full potential is now $\phi(r)=\phi_{{\rm{N}}}(r)+\phi_{{\rm{OU}}}(r)$. By the same arguments as above we find the coupling
\begin{equation}
\sigma\,=\,G_{{\rm{N}}}\,\frac{2}{2\,+\,\sqrt{\pi}}\sqrt{M\,M_{0}}\quad.
\label{OU4}\end{equation}

\subsection{Generalization of the Brownian process}
Now we consider a classical drift field $\mathbf{a}$ that is somewhat more general. The Fokker-Planck equation of interest is given by
\begin{equation}
r_{0}^{m}\,\partial_{r}\,r^{n}\,\phi_{{\rm{B}}}(r)\,+\,\frac{1}{r^{2}}\partial_{r}\,r^{2}\,\partial_{r}\,\phi_{{\rm{B}}}(r)+\,\rho\,=\,0\quad.
\label{GBp1}\end{equation}
For $n=0$ we obtain the ordinary Brownian motion, for $n=1$ we obtain the Ornstein-Uhlenbeck process. A choice for the power $0\leq n \leq 1$ allows us to study scenarios that deviate from ordinary Brownian motion but do not suffer from the Gaussian decay of the stochastic correction of the potential that is present in the case of the Ornstein-Uhlenbeck process, see Eq. (\ref{OU3}). Furthermore, we obtain an additional parameter that allows for a more advanced fitting to data sets. To illustrate the fidelity of our generalization we shall now treat a system that is described by
\begin{equation}
r_{0}^{-7/6}\,\partial_{r}\,r^{1/6}\,\phi_{{\rm{B}}}(r)\,+\,\frac{1}{r^{2}}\partial_{r}\,r^{2}\,\partial_{r}\,\phi_{{\rm{B}}}(r)+\,\rho\,=\,0\quad.
\label{GBp2}\end{equation}
The relevant part of the potential is given by
\begin{equation}
\phi_{{\rm{B}}}(r)\,=\,\sigma\,\left(\frac{7}{6}\right)^{6/7}\,\left(\frac{r_{0}}{r}\right)\,_{1}F_{1}\left[-\,\frac{5}{7},\,\frac{1}{7},\,-\,\frac{6}{7}\left(\frac{r}{r_{0}}\right)^{7/6}\right]\quad,
\label{GBp3}\end{equation}
where $_{1}F_{1}$ is Kummer's confluent hypergeometric function.

In the present case the coupling to the Newtonian has to be carried out by a renormalization of the gravitational constant that is given by $G_{{\rm{d}}}\,=\,G_{{\rm{N}}}\,\left(1\,-\,g\right)$. Note that the sign of $g$ is now negative. Consequently, we find the potential
\begin{equation}
\phi(r)\,=\,-\,\frac{G_{{\rm{N}}}\,M\,-\,G_{{\rm{N}}}\,\sqrt{M\,M_{0}}}{r}\,-\,\frac{G_{{\rm{N}}}\,\sqrt{M\,M_{0}}}{r}\,_{1}F_{1}\left[-\,\frac{5}{7},\,\frac{1}{7},\,-\,\frac{6}{7}\left(\frac{r}{r_{0}}\right)^{7/6}\right]\quad.
\label{GBp4}\end{equation}
We will refer to this potential as fractional-power potential. We assume that there is no deeper meaning in the different ansatz for the renormalized constant of gravity, it is just a mathematical procedure in order to obtain Newtonian gravity for $r\rightarrow 0$\,.

\subsection{Discussion}
\begin{figure}[t!]\centering\vspace{0.cm}
\rotatebox{0.0}{\scalebox{1.06}{\includegraphics{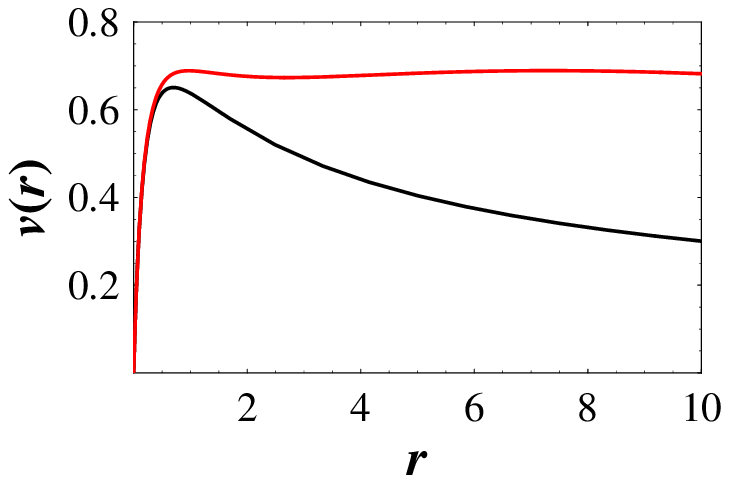}}}
\rotatebox{0.0}{\scalebox{1.06}{\includegraphics{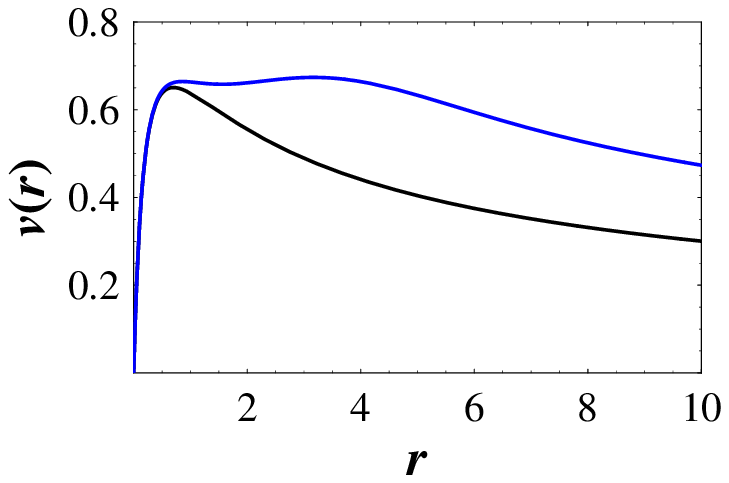}}}
\caption{\footnotesize{$G=1$\,, $M=1$\,, $\beta=1$\,. Left: Kepler-Newton velocity (black), Brownian-Moffat velocity (red). Parameters: $R_{c}=0.35$\,, $ r_{0}=5.5$\,, $M_{0}=53$\,. Right: Kepler-Newton velocity (black), Ornstein-Uhlenbeck velocity (red). Parameters: $R_{c}=0.35$\,, $r_{0}=\sqrt{3.5}$\,, $M_{0}=9$\,.}}
\label{figcomparison}\end{figure}

In Fig. (\ref{figcomparison}) we compare the rotation-curves that follow from the Brownian-Moffat model and the Ornstein-Uhlenbeck model. We easily see that the Brownian-Moffat model is able to produce flatness of the tail of $v(r)$ over a large distance, as it was reported in \cite{Brownstein1, Brownstein2, Moffat2}. For short to medium distances the Ornstein-Uhlenbeck model can also produce flatness, but obviously fails for larger distances. 

\begin{figure}[t!]\centering\vspace{0.cm}
\rotatebox{0.0}{\scalebox{1.06}{\includegraphics{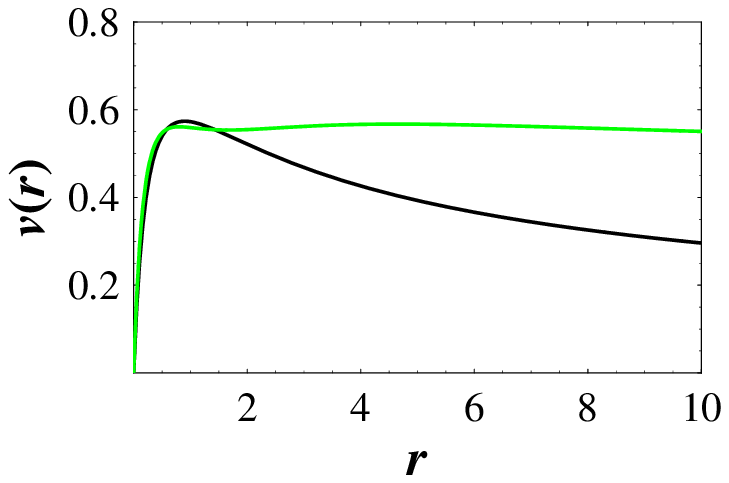}}}
\rotatebox{0.0}{\scalebox{1.05}{\includegraphics{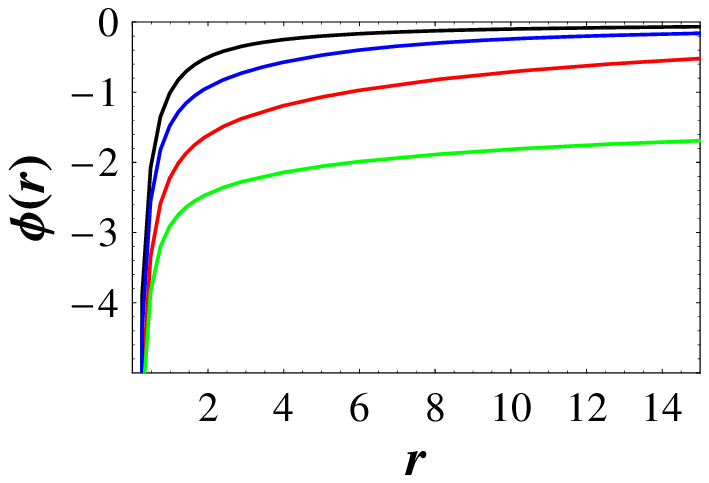}}}
\caption{\footnotesize{$G=1$\,, $M=1$\,, $\beta=1$\,. Left: Kepler-Newton velocity (black), fractional-power velocity (green). Parameters: The Kepler-Newton core is $R_{c}=0.45$\,, the parameters of the fractional-power system are $R_{c}=0.36$\,, $ r_{0}=1.043$\,, $M_{0}=0.26$\,. Right: Newtonian potential (black), Brownian-Moffat potential (red), Ornstein-Uhlenbeck potential (blue), fractional-power potential (green). For the parameters see Fig. (\ref{figcomparison}).}}
\label{figcomparison1}\end{figure}

By Fig. (\ref{figcomparison1}) we deduce that the Ornstein-Uhlenbeck process (blue) generates more attraction than the Newtonian in the regime where the cross-over from the tail to the center of the potential takes place. However, the augmented depth, by Fig. (\ref{figcomparison}), is not deep enough to generate flatness over a larger distance. The explanation here is that the Ornstein-Uhlenbeck correction decays like a Gaussian. This corresponds with the Brownian part of the trajectory, which decays like an exponential, such that the overall distortion of Newtonian gravity may be regarded to be moderate. Contrary to this, the Brownian-Moffat potential (red) is heavily attractive, both in it's tail and it's center. The augmented depth is deep enough to generate a flat rotation-curve also on larger distances. The heavy attractiveness corresponds to the drift-motion of the Brownian trajectory towards the center of gravity. The augmented depth generates a confinement with asymptotic freedom. However, we want to remark that the fact that the Ornstein-Uhlenbeck process cannot model flatness on larger distances does not outrule this process from a principal point of view. There very well may be rotation-curves that show more similarity with the Ornstein-Uhlenbeck process than with ordinary Brownian motion.

By Fig. (\ref{figcomparison1}) we see that fractional powers for the classical drift-field $\mathbf{a}$ can remove the fast decay of the Ornstein-Uhlenbeck model. Furthermore, we encounter an interesting effect that is not present in the Brownian-Moffat model and the Ornstein-Uhlenbeck model. The rotation-curves of the Kepler-Newton system and the fractional-power system must match for $r\rightarrow 0$\,. This but can only be achieved for different core-radii. In our example the Kepler-Newton system has a core-radius $R_{{\rm{c}}}=0.45$\,, while the fractional-power system has a core-radius $R_{{\rm{c}}}=0.36$\,. From this we can deduce two cases. If we suppose that the total extension of the Kepler-Newton system and the fractional-power system is equal, the total mass of both systems differs. The fractional-power system has slightly more total mass than the Kepler-Newton system. If we but suppose that the total mass of both systems is equal then the total extension differs. The Kepler-Newton system then is slightly larger than the fractional-power system. A match on data thus may show if a mass-defect is present in galaxies, or that the core is more condensed than it is assumed to be.

By the right-hand-sight of Fig. (\ref{figcomparison1}) we see that the fractional-power system generates a potential that is far more attractive than the Brownian-Moffat model. Asymptotic freedom but is guaranteed by $\lim_{r\rightarrow 0}\phi(r)=0$\,.

\section{Phase-diagrams}
In this section we discuss the properties of the parametric co-dimension $\{R_{{\rm{c}}},\,r_{0}, M_{0}\}$ that is provided by the above models. For simplicity we shall restrict ourselves on the Brownian-Moffat model Eq. (\ref{MSTG10}), since we expect that the parametric properties of the other models are not essentially different.

The starting-point here is the fact that the total angular momentum of the rotating system is conserved. Furthermore, we know that the rotational velocity is a velocity in azimuthal direction, thus $v_{\varphi}=v$. Consequently, the density of the angular momentum in $z$-direction is given by
\begin{equation}
l_{z}(r,\,R_{{\rm{c}}},\,r_{0},\,M_{0})\,=\,r\,M(r,\,R_{{\rm{c}}})\,v_{\varphi}(r,\,R_{{\rm{c}}},\,r_{0},\,M_{0})\quad.
\label{Angular1}\end{equation}
The total angular momentum is thus given by
\begin{equation}
L_{z}(R_{{\rm{c}}},\,r_{0},\,M_{0},\,r_{{\rm{max}}})\,=\,4\,\pi\,\int_{0}^{r_{{\rm{max}}}}dr\,r^{2}\,l_{z}(r,\,R_{{\rm{c}}},\,r_{0},\,M_{0})\quad,
\label{Angular2}\end{equation}
where $r_{{\rm{max}}}$ denotes the total extension of the system. Here we set $r_{{\rm{max}}}=10$\,. For a fixed mass-scale $M_{0}$ and a fixed angular momentum $L_{z}$ we can establish a relation between the core-radius $R_{{\rm{c}}}$ and the range or likewise correlation-length $r_{0}$. This relation may be understood as a phase-diagram.

\begin{figure}[t!]\centering\vspace{0.cm}
\rotatebox{0.0}{\scalebox{1.06}{\includegraphics{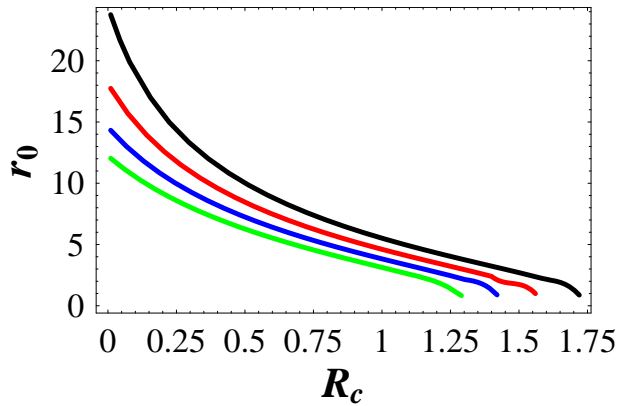}}}
\caption{\footnotesize{Phase-diagram for conserved angular momentum of the Brownian-Moffat model. $G=1$\,, $M=1$\,, $\beta=1$\,, $M_{0}=53$\,,  $r_{{\rm{max}}}=10$\,. $L_{z}=13,000$ (black), $L_{z}=14,000$ (red), $L_{z}=15,000$ (blue), $L_{z}=16,000$ (green). }}
\label{figangularmomentum1}\end{figure}

In Fig. (\ref{figangularmomentum1}) we have drawn four phase-lines which connect the core-radius $R_{{\rm{c}}}$ and the correlation length $r_{0}$ for conserved angular momenta. When we recall that the potential converges towards the Newtonian for growing correlation-length and vanishing core, $\phi(r,\, R_{{\rm{c}}}\rightarrow 0,\,r_{0}\rightarrow\infty)=\phi_{{\rm{N}}}(r)$, then we may understand the phase-lines as condensation-curves for a gravitating cloud. We deduce that for a fixed angular momentum $L_{z}$ and a fixed total extension $r_{{\rm{max}}}$ the possible extension of the core is limited from above. The state of a highly extended core may be understood as the initial state. Driven by it's internal dynamics the system condenses along it's phase-line until the Kepler-Newton state is reached. We emphasize that for the final state $R_{{\rm{c}}}=0$ is always possible, while the final value of the correlation-length $r_{0}$ crucially depends on the value of the total angular momentum. We observe that for higher angular momenta the parameter-space is smaller than for lower angular momenta. Thus, the lower the total angular momentum, the more extended may the initial core be and the more properly can a final Kepler-Newton state be reached. The black line in Fig. (\ref{figangularmomentum1}) illustrates a case where the final Kepler-Newton state is almost completely realized, but this does not fully hold for the green line. This case illustrates a scenario where regions with a large distance to the center still may show rotational anomalies. For even higher angular momenta we may conclude that a proper Kepler-Newton state is not possible. In summary, from Fig. (\ref{figangularmomentum1}) we see that a phase-line can be drawn for each galaxy, depending on it's respective parameters. From the position on it's phase-line information about the internal state can be read off immediately.

The same may hold for proto-planetary disks. Inspection of the respective phase-line can show immediately if the final state of a Kepler-Newton solar-system can be reached. If the total angular-momentum is too high it is likely that this is impossible.

However, all what has been said so far only holds in the case of a smooth evolution. Chaos or catastrophic events but of course can cause drastic modifications of the system. This then leads either to a changed position on the phase-line or to the bifurcation of a new phase-line.

\begin{figure}[t!]\centering\vspace{0.cm}
\rotatebox{0.0}{\scalebox{1.06}{\includegraphics{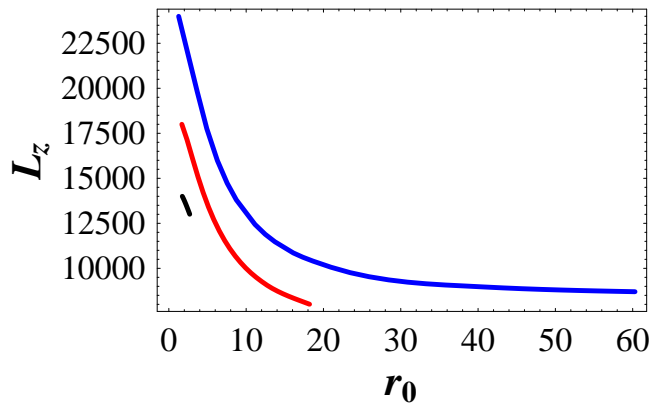}}}
\caption{\footnotesize{Phase-diagram of existence of the Brownian-Moffat model. $G=1$\,, $M=1$\,, $\beta=1$\,, $M_{0}=53$,\,$r_{{\rm{max}}}=10$\,. Existence-curve for $R_{{\rm{c}}}=1.5$ (black), existence-curve for $R_{{\rm{c}}}=1.0$ (red), existence-curve for $R_{{\rm{c}}}=0.5$ (blue).}}
\label{figangularmomentum2}\end{figure}

By Fig. (\ref{figangularmomentum2}) we illustrate the relation between the total angular momentum $L_{z}$ and the correlation-length $r_{0}$. This relation defines phase-lines of existence. Again we find that for high angular momenta the correlation-length remains comparably small. The black line is of considerable interest. By inspection we see that a system with core-radius $R_{{\rm{c}}}=1.5$, mass-scale $M_{0}=53$ and total extension $r_{{\rm{max}}}=10$ can only exist in a narrow window of angular momentum $13,000\leq L_{z}\leq 14,000$\,. This corresponds to the black and red phase-lines in Fig. (\ref{figangularmomentum1}). In the frame of our above picture of condensation this implies that any rotating system with extension $r_{{\rm{max}}}=10$ and mass-scale $M_{0}=53$ should have an angular momentum that is located in the interval $13,000\leq L_{z}\leq 14,000$\,. Regardless of how condensed the core of such a system is, we may assume that it's initial value should approximately have been $R_{{\rm{c}}}=1.5$\,. If now a system with  $r_{{\rm{max}}}=10$ and mass-scale $M_{0}=53$ has an angular momentum $L_{z}=16,000$, then we may conclude that it's initial core-radius should have been $R_{{\rm{c}}}=1$\,. This corresponds to the red phase-line in Fig. (\ref{figangularmomentum2}). Consequently, it is possible to identify the likely initial state of a system by knowledge of the fundamental parameters $\{L_{z},\, M_{0},\,r_{{\rm{max}}}\}$. As above, this only but holds for a smooth evolution. Chaos or catastrophic events can change the fundamental parameters of a system. Thus, by analyzing the phase-lines of existence we can gain knowledge about the likely evolution of a system.

As an example we take a galaxy with the fundamental parameters $\{L_{z}=12,000\,, M_{0}=53,\,r_{{\rm{max}}}=10\}$, that is located on the blue line in  Fig. (\ref{figangularmomentum2}). This galaxy also exists on the red line, but does not match with the small black line. From this two cases follow. The trivial case is that this galaxy had an initial core of radius $R_{{\rm{c}}}=1$\,, and we are done. The non-trivial case, however, is that by some event one, two or three of the fundamental parameters have been changed. This is a necessary conclusion in order to explain the angular momentum, which does not match with the small black line.

Finally, we observe that with shrinking core the region of existence grows. Systems with a point-like or almost point-like core, blue line in Fig. (\ref{figangularmomentum2}), exist for any value of angular momentum and correlation-length, as it was to be expected. Consequently, the Kepler-Newton state is a universal state, while the initial state of a gravitating cloud is special and restricted by a certain range of parameters.

\section{Conclusion}
We successfully related the weak-field limit of MSTG and the anomalous rotation of galaxies to Brownian motion and diffusion processes. We have shown that a stochastic distortion can lead to potential-tails that are far more attractive than the Newtonian. This phenomenon is qualified to make the assumption of dark matter unnecessary. The parameters that enter the full potential are clearly defined by the core-radius $R_{{\rm{c}}}$, a mass-scale $M_{0}$ and the range $r_{0}$, which may be regarded as fundamental properties of the galaxy under consideration. The resulting rotation-curves of course depend on the chosen model for the mass.

Furthermore, we have introduced phase-diagrams which connect the fundamental parameters of the system. For a fixed angular momentum we have found phase-lines that may explain the condensation-process of gravitating clouds towards the final Kepler-Newton state. For a fixed value of the core-radius we were able to calculate phase-lines that describe regions of existence for gravitating objects with respect to their angular momentum and their range. We found that for systems with highly extended core there exists only a small window with respect to angular momentum and correlation-length where a gravitating object can exist. As the core condenses this window grows until the whole possible range of the parametric co-dimension is covered. The initial restriction for extended cores allows to determine the likely initial state of a galaxy, while the phase-diagram for conserved angular momentum allows to observe the likely evolution towards a final Kepler-Newton state. We assume that these results also may hold for proto-planetary disks.

As galaxies are classical many-body systems it is likely that dynamics which deviates from Newtonian behaviour on large scales is related to Brownian processes. Thus, we assume that our results are not just by chance.

So far we have used Gaussian noise to model the stochastic correction. A possible generalization may be given by the introduction of L\'evian noise. This, however, shall be left for further work.


\begin{thebibliography}{99}

\bibitem{Bekenstein}
Bekenstein J. D.: Relativistic gravitation theory for the MOND paradigm,  Phys. Rev. D 70, 083509, (2004), arXiv:astro-ph/0403694, (2004).

\bibitem{Blanchet1}
Blanchet L.: Gravitational polarization and the phenomenology of MOND, Class. Quantum Grav. 24, 3529–3539, (2007), arXiv:astro-ph/0605637, (2007).

\bibitem{Blanchet2}
Blanchet L., Heisenberg L.: Dipolar Dark Matter with Massive Bigravity, arXiv:1505.05146, (2015).

\bibitem{Brownstein1}
Brownstein J. R., Moffat J. W.: Galaxy Rotation Curves Without Non-Baryonic Dark Matter, Astrophys. J. 636, 721, (2006),  arXiv:astro-ph/0506370, (2005).

\bibitem{Brownstein2}
Brownstein J. R., Moffat J. W.: The Bullet Cluster 1E0657-558 evidence shows Modified Gravity in the absence of dark matter, Mon. Not. Roy. Astron. Soc. 382, 29, (2007), arXiv:astro-ph/0702146, (2007).

\bibitem{Deffayet}
Deffayet C., Esposito-Farese G., Woodard R. P.: Nonlocal metric formulations of MOND with sufficient lensing, Phys.Rev. D {\bfseries{84}} 124054, (2011),  	arXiv:1106.4984, (2011).

\bibitem{Hossenfelder}
Hossenfelder S., Mistele T.: The Redshift-Dependence of Radial Acceleration: Modified Gravity versus Particle Dark Matter, arXiv:1803.08683v1, (2018).

\bibitem{Kaplinghat}
Kaplinghat M., Turner M.: How Cold Dark Matter Theory Explains Milgrom's Law, Astrophys. J. 569, L19, (2002),  arXiv:astro-ph/0107284, (2002).

\bibitem{Milgrom1}
Milgrom M.: A modification of the Newtonian dynamics as a possible alternative to the hidden mass hypothesis, Astrophys. J. 270, 365-370, (1983).

\bibitem{Milgrom2}
Milgrom M.: A modification of the Newtonian dynamics - Implications for galaxies, Astrophys. J. 270, 371–389, (1983).

\bibitem{Milgrom3}
Milgrom M.: MOND Laws of Galaxy Dynamics, arXiv:1212.2568, (2013).

\bibitem{Moffat1}
Moffat J. W.: Nonsymmetric Gravitational Theory, Phys. Lett. B 355, 447, (1995), arXiv:gr-qc/9411006, (1995).

\bibitem{Moffat2}
Moffat J. W.: Gravitational Theory, Galaxy Rotation Curves and Cosmology without Dark Matter, JCAP 0505, (2005), arXiv:astro-ph/0412195, (2005).

\bibitem{Ostriker}
Ostriker J. P., Steinhardt P. J.: The observational case for a low-density Universe with a non-zero cosmological constant, Nature 377, 600, (1995).

\bibitem{Rubin1}
Rubin V., Ford Jr., Kent W.: Rotation of the Andromeda Nebula from a Spectroscopic Survey of Emission Regions, Astrophys. J. 159, 379ff., (1970).

\bibitem{Rubin2}
Rubin V., Roberts M. S., Graham J. A., Ford Jr., Kent W., Thonnard N.: Motion of the Galaxy and the Local Group determined from the Velocity Anisotropy of distant SC 1 Galaxies, I and II, Astrophys. J. 81, 687 and 719ff., (1976).

\bibitem{Sanders}
Sanders R. H.: A tensor-vector-scaler framework for modified dynamics and cosmic dark matter, arXiv:astro-ph/0502222, (2005).

\bibitem{Verlinde}
Verlinde E. P.: Emergent Gravity and the Dark Universe, SciPost. Phys. 2, 016, (2017), arXiv:1611.02269, (2017). 




\end{thebibliography}
\end{document}